\documentclass[prb,twocolumn,superscriptaddress]{revtex4}
\usepackage{graphicx}

\begin{document}

\bibliographystyle{revtex}
\title{Anisotropic Vortex Pinning in Superconductors
       with a Square Array of Rectangular Submicron Holes}

\author{L. Van Look}
\affiliation{Laboratorium voor Vaste-Stoffysica en Magnetisme,\\
K. U. Leuven\\ Celestijnenlaan 200 D, B-3001 Leuven, Belgium}

\author{B. Y. Zhu}
\affiliation{Laboratorium voor Vaste-Stoffysica en Magnetisme,\\
K. U. Leuven\\ Celestijnenlaan 200 D, B-3001 Leuven, Belgium}

\author{R. Jonckheere}
\affiliation{Inter-University Micro-Electronics Center (Imec vzw),
Kapeldreef 75, B-3001 Leuven, Belgium}

\author{B.~R.~Zhao}
\affiliation{National Laboratory for Superconductivity, Institute
of Physics and Center for Condensed Matter Physics, Chinese
Academy of Sciences, Beijing 100080, China}

\author{Z.~X.~Zhao}
\affiliation{National Laboratory for Superconductivity, Institute
of Physics and Center for Condensed Matter Physics, Chinese
Academy of Sciences, Beijing 100080, China}

\author{V.~V.~Moshchalkov}
\affiliation{Laboratorium voor Vaste-Stoffysica en Magnetisme,\\
K. U. Leuven\\ Celestijnenlaan 200 D, B-3001 Leuven, Belgium}

\date{\today}

\begin{abstract}

We investigate vortex pinning in thin superconducting films with a
square array of rectangular submicron holes (\textit{"antidots"}).
Two types of antidots are considered: antidots fully perforating
the superconducting film, and \textit{"blind antidots"}, holes
that perforate the film only up to a certain depth. In both
systems, we observe a distinct anisotropy in the pinning
properties, reflected in the critical current $I_c$, depending on
the direction of the applied electrical current: parallel to the
long side of the antidots or perpendicular to it. Although the
mechanism responsible for the effect is very different in the two
systems, they both show a higher critical current and a sharper
IV-transition when the current is applied along the long side of
the rectangular antidots.

\end{abstract}

\pacs{PACS numbers: 74.76.Db, 74.60.Ge, 74.25.Dw, 74.60.Jg,
74.25.Fy}

\maketitle


\section{Introduction}

Type-II superconductors (SC's) with nano-engineered artificial
pinning arrays are good candidates to study the fundamentals of
vortex pinning, since, within the limits of the lithographic
process used for their fabrication, the pinning centers can be
shaped and positioned in the SC at will. Often used systems in
that respect are, for example, periodic arrays of submicron holes
(antidots)\cite{hebard77ieee,baert95prl,vvm96prb,vvm98prb,castellanos:97apl}
or magnetic dots, placed underneath or on top of the SC
film\cite{vanbael99prb,mvb:00prl,martin97prl,morgan:98prl,Otani93jmagnmagnmater}.

In SC's with a periodic pinning array (PPA), so-called matching
effects\cite{baert95prl} occur, at specific magnetic fields
generating a number of vortices which ``matches" the number of
available pinning sites. At these integer $H_n$ and fractional
$H_{p/q}$ matching fields, the vortices form regular geometrical
patterns, commensurate with the pinning array. This strongly
reduces the vortex mobility and consequently increases the
critical current $I_c$. These commensurability effects have been
recently intensively studied for square or triangular arrays of
antidots\cite{baert95prl,vvm96prb,vvm98prb} or magnetic
dots\cite{vanbael99prb,martin97prl,Otani93jmagnmagnmater}.

In case of a square PPA, the equivalence between the two important
in-plane directions, [10] and [01] ($x$- and $y$-axis), is
conserved if square or cylindrical antidots (or out-of-plane
magnetic dots) are used. However, as shown by recent numerical
simulations, anisotropic pinning properties can be introduced in
an otherwise isotropic superconducting film by using for example a
\textit{rectangular array} of \textit{isotropic} pinning
sites~\cite{reichhardt1100losalamos}. In the present work we
study, by means of experiments and numerical simulations, the
anisotropy in the pinning properties of a SC film caused by a
\textit{square array} of \textit{rectangular antidots} and
\textit{blind antidots}.

This paper is organized as follows. First, we consider rectangular
antidots that perforate the superconductor completely. For this
system, we present experimental results and use numerical
simulations to gain more insight in the mechanisms responsible for
the observed anisotropy in the vortex mobility and dynamics. In a
second part, we investigate rectangular antidots that do not
perforate the superconducting film completely ("\textit{blind
antidots}"). Here, we will present only numerical simulations.
Experiments on a square array of blind rectangular antidots have,
to our knowledge, not been performed sofar.

\section{Square array of rectangular antidots}
\subsection{Electrical transport measurements}
\subsubsection{Experimental details}

\begin{figure}[ht]
  \centering
  \includegraphics*[scale=0.6]{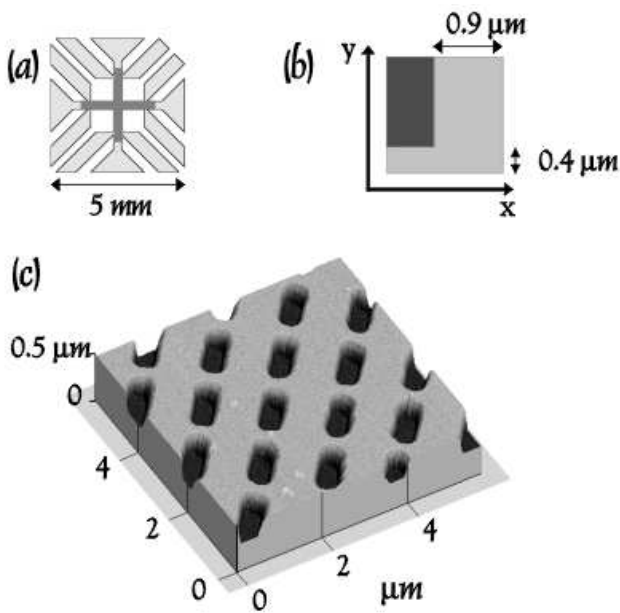}
  \caption{Layout of the Pb film with a square array of rectangular
  antidots. (a) Cross-shaped geometry of the sample to allow for
  transport measurements in the $x$- and $y$-direction. (b) Schematic representation
  of a unit cell (1.5~$\times$~1.5~$\mu$m$^2$) of the antidot array. (c) Atomic force micrograph of
  a 6~$\times$~6~$\mu$m$^2$ area of the antidot lattice.}
  \label{fig:layout}
\end{figure}

We patterned a SC Pb film in a $5 \times 5 $ mm$^2$ cross-shaped
geometry (see Fig. \ref{fig:layout}(a)) to allow electrical
transport measurements in two perpendicular current directions.
The central part of the cross consists of two 300~$\mu$m wide
strips containing the square array of rectangular antidots (see
dark gray area in Fig. \ref{fig:layout}(a)). In both strips, the
long side of the antidots points in the $y$-direction. This
pattern was prepared by electron-beam lithography in a polymethyl
metacrylate/methyl metacrylate (PMMA/MMA) resist bilayer covering
the SiO$_2$ substrate. A Ge(20~\AA)/Pb(1500~\AA)/Ge(300~\AA) film
was then electron-beam evaporated onto this mask while keeping the
substrate at 77 K. After a liftoff process in warm acetone, the
structure was covered with a thick insulating Ge(1000~\AA) layer
for protection against oxidation.

Figure \ref{fig:layout}(c) shows an atomic force microscopy (AFM)
topograph of a $6 \times 6~\mu$m$^2$ area of the square antidot
lattice with a period of 1.5 $\mu$m. We see that the antidots have
a rectangular shape $(0.6~\times~1.1 ~\mu $m$^2$) with rounded
corners. As shown in the schematic representation of a unit cell
of the array in Fig.~\ref{fig:layout}(b), the superconducting
paths between the antidots are 0.4 $\mu$m and 0.9 $\mu$m wide, for
the $x$- and $y$-direction, respectively. The rms roughness value
of the Ge/Pb/Ge film, in between the antidots, is less than
15~\AA~ on a 1~$\mu$m$^2$ area.

The transport measurements were performed in a $^4$He cryostat
equipped with a 9 T superconducting magnet. The magnetic field was
applied perpendicular to the film surface. Four-point resistivity
measurements were carried out using an ac bridge at a frequency of
19 Hz and a current of 10 $\mu$A. We obtained the superconducting
critical temperature ($T_c$ = 7.26~K) using a resistance criterion
of 10 \% of the normal state resistance $R_n$. The ratio of the
$R_n$ values for the $x$- and $y$-direction is
 \[ R_n^{x} \slash R_n^{y}
 = 1.06 ~\Omega \slash 0.46 ~\Omega=2.3, \] which is in excellent
 agreement with the value 2.25 that can be expected from geometrical considerations.
To determine the superconducting coherence length $\xi(0)$, we
measured the linear $T_c(H)$ phase boundary of a co-evaporated
reference film without any in-plane nanostructuring. From the
$T_c(H)$ slope, and\cite{tinkhambook}
\begin{equation}
\mu_0 H_{c2}=\frac{\Phi_0}{2 \pi \xi(T)^2}=\frac{\Phi_0}{2 \pi
\xi(0)^2}\left(1-\frac{T}{T_c}\right) \, ,
\end{equation}
we find $\xi(0)$ = 51~nm. Using the dirty limit ($\ell < \xi_0$)
expression $\xi(0)=0.865\sqrt{\xi_0 \ell}$ and the BCS coherence
length for Pb, $\xi_0$= 83 nm\cite{dgabook}, we determined the
elastic mean free path $\ell=42$~nm. We can derive the penetration
depth $\lambda(0)=34$~nm by means of the dirty limit expression
\begin{math}\lambda(0)=0.66 \lambda_L \sqrt{\xi_0 \slash
\ell}\end{math}, using $\lambda_L$=37~nm as the London penetration
depth\cite{dgabook}. Taking into account that the presence of
antidots in a superconducting film has the tendency to increase
the penetration depth in the following way\cite{wahl:95physicac}:
\begin{equation}
\Lambda(0)=\frac{\lambda(0)}{\sqrt{1-2\frac{S_a}{S_t}}}=53~
\mathrm{nm} \, ,
\end{equation}
where $S_a$ is the area occupied by the antidots and $S_t$ is the
total area of the film, we obtain a Ginzburg-Landau (GL) parameter
\begin{math} \kappa = \Lambda(0) \slash \xi(0) \approx 1 > 1 \slash \sqrt{2} \end{math}.
We therefore conclude that the patterned film with the array of
rectangular antidots is a type-II superconductor.

\subsubsection{Results}

\begin{figure}[ht]
  \centering
  \includegraphics*[scale=0.75]{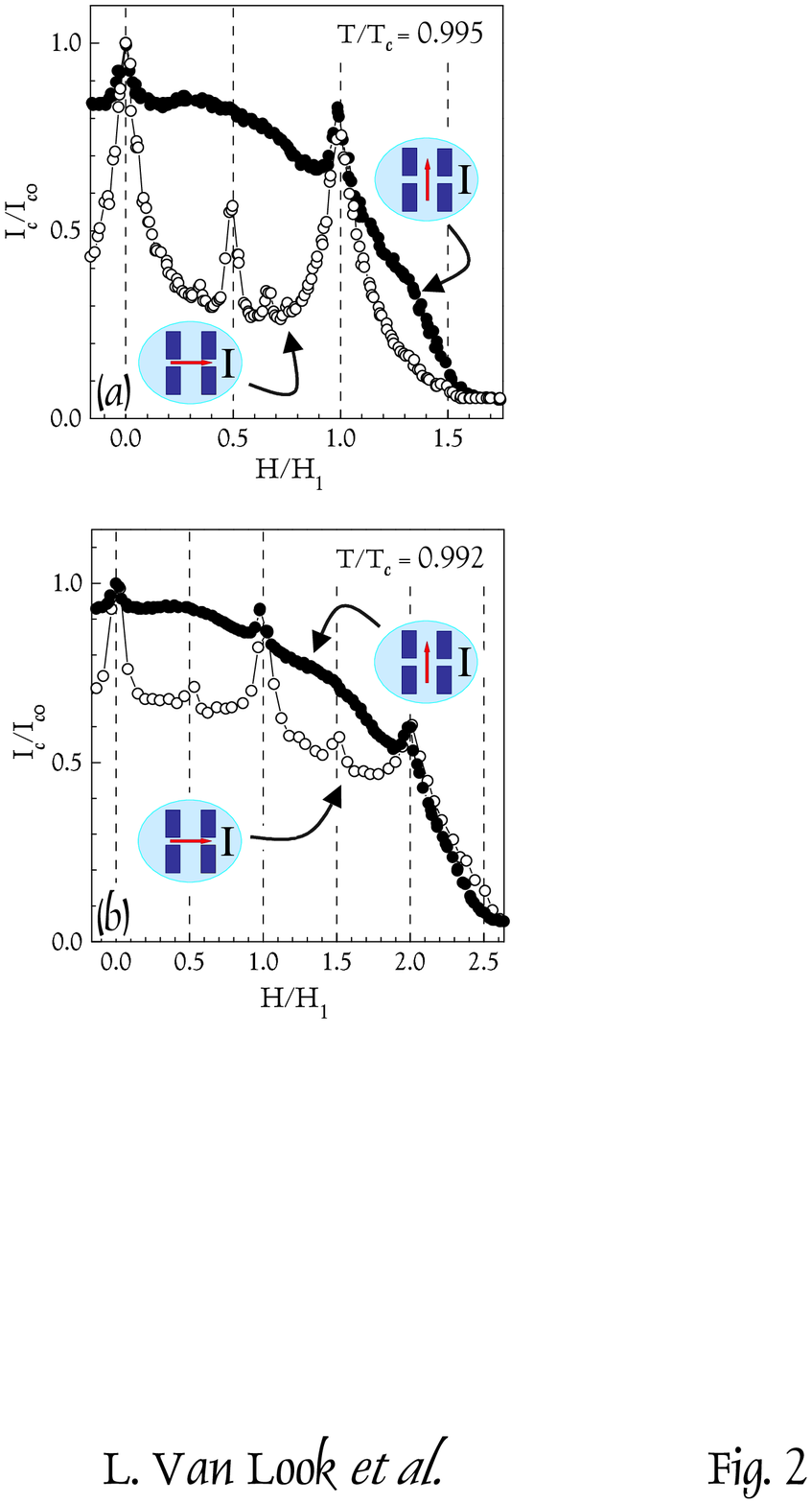}
  \caption{Normalized critical currents $I_{cx}$ and $I_{cy}$ as a
  function of normalized magnetic field $H/H_1$ for the sample shown in Fig.~\protect\ref{fig:layout} measured
  with a current in the $x$- (open symbols) and the $y$-direction (filled
  symbols), respectively. The $I_c(H)$ data are presented for two temperatures: (a) $T/T_c$ =
  0.995 and (b) $T/T_c$ = 0.992.}
  \label{fig:icb}
\end{figure}

In Fig.~\ref{fig:icb}, we show the critical current versus field
curves $I_c(H)$, normalized to its value at zero field, $I_{co}
\equiv I_c(H = 0)$, at two temperatures ($T/T_c$ = 0.995 in (a)
and 0.992 in (b)). We have used a voltage criterion of
\begin{math} V_{crit}=100~R_n~\mu
\end{math}V/$\Omega$, in order to be able
to make a comparison between the two current directions with
different normal state resistances $R_n^x$ and $R_n^y$. The
critical current density at zero field measured along the
$x$-direction is $I_{co}^x=~4.4~\cdot~10^{7}
\frac{\mathrm{A}}{\mathrm{m}^2}$ at $T/T_c=0.995$ and $I_{co}^x=
9.7~\cdot~10^{7} \frac{\mathrm{A}}{\mathrm{m}^2}$ at
$T/T_c=0.992$. For a current along the $y$-direction, these values
are $I_{co}^y=1.3~\cdot~10^{8} \frac{\mathrm{A}}{\mathrm{m}^2}$ at
$T/T_c=0.995$ and $I_{co}^y= 2.0~\cdot~10^{8}
\frac{\mathrm{A}}{\mathrm{m}^2}$ at $T/T_c=0.992$. The field axis
is given in units of the first matching field $H_1$, the field at
which the density of $\phi_0$-vortices equals the density of
antidots:
\begin{math} \mu_0H_1=\Phi_0\slash d^2=9.2~\text{Oe}
\end{math}, with $d=1.5~\mu$m the period of the antidot lattice
and $\phi_0=\frac{h}{2e}$ the superconducting flux quantum.

Due to the presence of the antidot lattice, both the $I_{cx}(H)$
and $I_{cy}(H)$ data show pronounced maxima at the integer
matching fields $H_1$ and $H_2$ up to \textit{the same critical
current value}.
The differences between the two current directions appear in the
field ranges between the integer matching fields. For those field
intervals, the critical current $I_{cy}(H)$ (filled symbols in
Fig. \ref{fig:icb}) is considerably enhanced compared to
$I_{cx}(H)$ (open symbols). This increase of $I_{cy}$ is
accompanied by the complete suppression of the rational matching
peaks at $H_{p/q}$ ($p$ and $q$ integers). For $I_{cx}(H)$, on the
other hand, the rational matching peaks at $H_{1/2}, H_{1/3},
H_{2/3}$ (panel (a)), and also at $3/2~H_1$ (panel (b)) are
clearly revealed.

\begin{figure}[b]
  \centering
  \includegraphics*[scale=1]{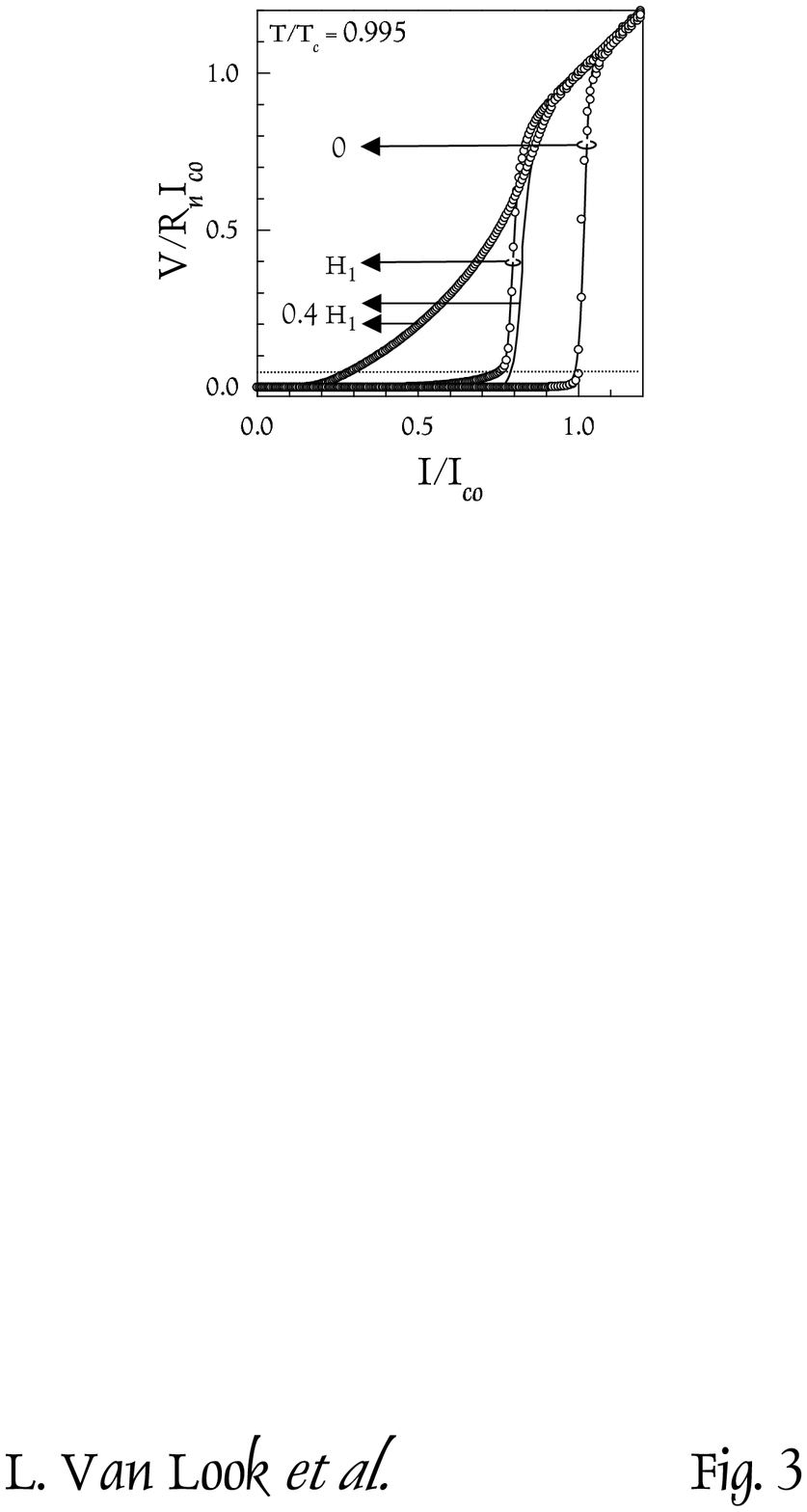}
  \caption{Normalized $V(I)$ curves at selected magnetic field values at the same
  temperature ($T/T_c$ = 0.995) as the measurements in Fig.~\protect\ref{fig:icb}(a).
  The full lines are the data for $I~\|~\mathbf{y}$, the open symbols for
  $I~\|~\mathbf{x}$. The horizontal dotted line indicates the voltage criterium
  used to define the critical current. For $H$~=~0 and $H~=~H_1$, the curves for the two current
  directions overlap almost completely.}
  \label{fig:iv}
\end{figure}

To investigate the origin of this qualitative difference in the
behavior of the critical currents along the two directions, we
have a closer look at the $V(I_x)$ (open symbols) and $V(I_y)$
(solid lines) curves for some selected field values at
$T/T_c=0.995$ (Fig.~\ref{fig:iv}). To make a comparison possible,
the voltage axis is rescaled to $R_nI_{co}$, and the current axis
to $I_{co}$. The dotted horizontal line indicates the voltage
criterium that was used to define the critical current shown in
Fig.~\ref{fig:icb}.

At $H = 0$ and $H = H_1$, the $V(I)$ transitions are very sharp
and independent of the current direction. The $V(I_x)$ and
$V(I_y)$ curves in Fig.~\ref{fig:iv} are therefore almost
indistinguishable. At $H=H_1$, the appearance of a low-voltage
tail in the $V(I)$ curve can be observed due to a small deviation
from the first matching field. These $V(I)$ curves have a shape
that is typical for a regular pinning
array\cite{conciseencyclopedia,martinoli78prb} as it was e.g. also
observed in SC films with a periodically modulated
thickness\cite{daldini75solidstatecommun}.


For $H = 0.4~H_1$, the current at which the film completely
reaches the normal state is the same for the two current
directions, but a large tail in the $V(I_x)$ curve (open symbols)
is present, indicating the dissipative motion of vortices in a
direction perpendicular to the current. Consequently, the $V(I_x)$
transition is very broad (transition width $\Delta I =
0.8~I_{co}$) compared to a much smaller broadening for $V(I_y)$
($\Delta I = 0.2~I_{co}$). This behavior is typical for every
magnetic field in between the matching fields.

\begin{figure}[ht]
  \centering
  \includegraphics*[scale=0.65]{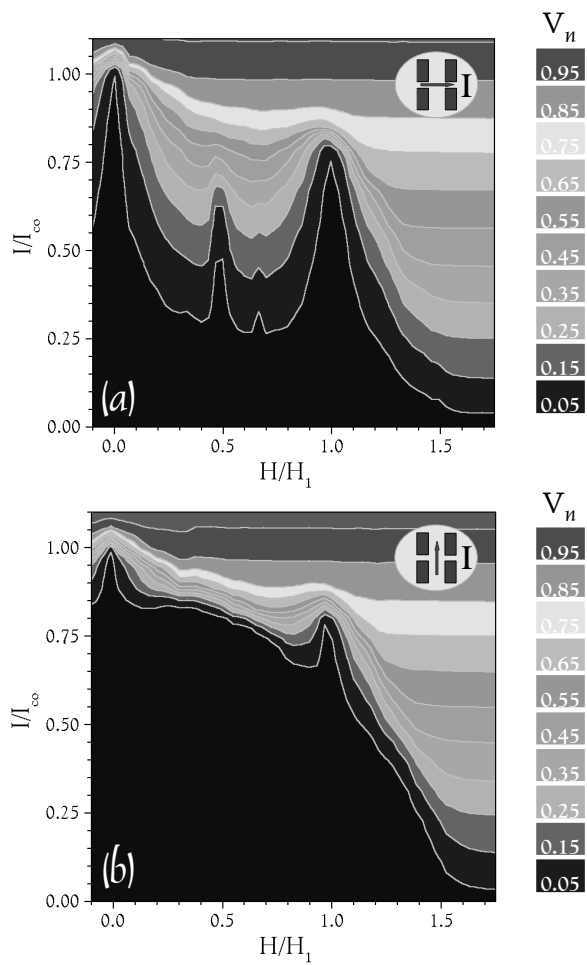}
  \caption{Contour plots (a) $V(I_x,H)$ and (b) $V(I_y,H)$ at
  $T/T_c$ = 0.995. The gray scale indicates the normalized voltage $V_n=\frac{V}{R_nI_{co}}$.
  The evolution of the $V(I)$ transition width as a function of
  magnetic field can easily be seen from this plot.}
  \label{fig:contourplot}
\end{figure}
The field dependence of the transition width $\Delta I$ can be
more adequately shown by presenting the $V(I_x,H)$ and $V(I_y,H)$
curves in a contour plot (Fig.~\ref{fig:contourplot}(a) and (b)).
We see that, at $H=0$ and at $H=H_1$, the $V(I)$ transitions are
equally sharp for both current directions. In between the integer
matching fields, a substantial broadening of the $V(I_x)$
transition takes place (Fig. \ref{fig:contourplot}(a)). At the
rational matching fields $H_{1/3},~H_{1/2},$ and $H_{2/3}$, the
transitions regain part of their sharpness. For $I~\|~\mathbf{y}$
(Fig. \ref{fig:contourplot}(b)), all $V(I_y)$ curves, both at and
in between the integer matching fields, have a sharp transition.

Summarizing the experimental observations, we have found a strong
anisotropy in the critical current $I_c(H)$ and the $V(I)$
characteristics of a film with a square array of rectangular
antidots. At the integer matching fields $H_n$, both the critical
current $I_c(H_n)$ and the $V(I,H_n)$ curve do not depend on the
direction of the applied current - $I~\|~\mathbf{x}$ or
$I~\|~\mathbf{y}$. However, in between the integer matching
fields, we find broad $V(I_x)$ transitions, a low $I_{cx}$, and
rational matching features in $I_{cx}(H)$. For $I~\|~\mathbf{y}$,
on the other hand, we see sharp $V(I_y)$ transitions at every
magnetic field, an $\it{overall}$ high $I_{cy}$, and no sign of
rational matching features in $I_{cy}(H)$.

\subsubsection{Discussion}
\label{sec:exp_discussion}

 The differences between $I_{cx}(H)$ and $I_{cy}(H)$ are
due to an anisotropy in the vortex mobility in the sample along
the two in-plane directions. Neglecting the thermal noise force,
the velocity $\mathbf{v}$ of a vortex in a PPA is determined by
the superposition of three forces: the vortex-vortex interaction
$\mathbf{F}_{VV}$, the vortex-antidot pinning force $\mathbf{F}_P$
and the Lorentz force $\mathbf{F}_L$, directed perpendicular to
the applied current. The critical current $I_c$ of the system is
proportional to the Lorentz force $\mathbf{F}_L$ which is needed
to induce a threshold average vortex velocity, and therefore a
certain critical voltage. To reach this voltage, a noticeable
fraction of the vortices must be depinned by the joint effect of
the Lorentz $\mathbf{F}_L$ and the vortex-vortex interaction
$\mathbf{F}_{VV}$ forces. The balance between these forces, and
consequently $I_c$, depends strongly on the applied magnetic
field, since the latter determines the density of vortices in the
sample.

When, for example, all pinning sites in a square PPA are occupied
by a vortex (first matching field $H_1$), the vortex lattice has
such a high symmetry, that all vortex-vortex interactions between
the vortices trapped in the pinning sites cancel out
($\mathbf{F}_{VV}=0$). Therefore, at this field, the Lorentz force
$\mathbf{F}_{L}$ which is needed for depinning, is only determined
by the pinning force. This implies that the critical current at
the first matching field $I_c(H_1)$ is a measure of the
\textit{single site pinning force} of the individual antidots.
Once the Lorentz force exceeds the pinning force, all vortices
leave their pinning site simultaneously, resulting in a sharp
$V(I)$ transition \cite{reichhardt0800losalamos}.

Since the $I_{cx}(H_1)$ and $I_{cy}(H_1)$ values and the
$V(I_x,H_1)$ and $V(I_y,H_1)$ curves (Figs.~\ref{fig:icb}
and~\ref{fig:iv}) coincide, we conclude that the pinning force
exerted by a rectangular antidot on a single pinned
$\phi_0$-vortex is \textit{isotropic} along the two symmetry-axes
of the rectangular antidot. This experimental observation is in
agreement with predictions by Buzdin and
Daumens\cite{buzdin98physicac} for an elliptic antidot, where the
pinning force is only expected to be anisotropic when the ellipse
is extremely elongated.

The observed anisotropy in pinning properties can therefore
\textit{not} be attributed to an anisotropic single site pinning
force of the rectangular antidots themselves, but rather to their
arrangement in the array. In other words, \textit{the observed
anisotropy in the pinning properties should be associated with an
anisotropic vortex-vortex interaction}. Indeed, since the SC
strands between the antidots, where the screening currents of the
vortices have to flow, are much thinner between the adjacent
antidots in the $y$-direction than in the $x$-direction, we expect
the $y$-component of the vortex-vortex interaction forces to be
considerably larger than their $x$-component.

To understand how the anisotropic vortex-vortex interaction can
give rise to the observed anisotropy in the pinning properties in
between the matching fields, we examine the role of the
vortex-vortex interaction in the depinning process of vortices in
a regular pinning array.

We have discussed earlier that at a matching field, the
vortex-vortex interaction vectors are canceled out and depinning
is governed by the single site pinning force. When the applied
field is now slightly detuned from a matching field (integer or
rational), the vortex lattice contains defects due to the
incommensurability of the vortex array and the PPA. The vortex
rows, parallel to $\mathbf{F}_L$, without such defects are called
``commensurate'' rows. In these commensurate vortex rows, the
component of $\mathbf{F}_{VV}$ parallel to $\mathbf{F}_L$ will be
zero for all vortices in the row. On the other hand, in vortex
rows containing defects, the vortex-vortex interaction forces will
not cancel out at all ($\mathbf{F}_{VV}\neq 0$). This
$\mathbf{F}_{VV}$ component in the direction of the Lorentz force
$\mathbf{F}_L$ will assist to the depinning of these
``incommensurate'' rows, i.e. depinning occurs when the sum of
this $\mathbf{F}_{VV}$ component and $\mathbf{F}_L$ overcomes the
single site pinning force. The commensurate rows stay pinned up to
a higher Lorentz force $\mathbf{F}_L$~\cite{reichhardt97prl}. If
the vortex-vortex interaction is sufficiently large, the critical
current $I_c(H)$ will therefore be substantially lower right
before and after the matching fields, leading to pronounced
matching peaks (integer and rational).

When the current is applied in the $x$-direction
($\mathbf{F}_L~\|~\mathbf{y}$), the $I_{cx}(H)$ curve shows
pronounced maxima at the rational matching fields $H_{1/2}$,
$H_{3/2}$, $H_{1/3}$ and $H_{2/3}$ and a much lower value in
between these rational matching fields (see Fig.~\ref{fig:icb}).
This is a clear sign of a strong vortex-vortex interaction
$\mathbf{F}_{VV}$ component along the direction of the Lorentz
force $\mathbf{F}_L$, in this case the $y$-direction.

When a current is applied in the $y$-direction
($\mathbf{F}_L~\|~\mathbf{x}$), we see a complete disappearance of
the rational matching peaks and, instead, an overall high critical
current $I_{cy}(H)$ (Fig.~\ref{fig:icb}). This indicates that the
$\mathbf{F}_{VV}$ components along the direction of the Lorentz
force (in this case the $x$-direction) are much weaker than in the
case $I~\|~\mathbf{x}$. The vortex-vortex interaction is
apparently not able to provide the long-range order needed to
generate the sparse geometrical patterns at the rational matching
fields $H_{p/q}$. This implies that no commensurate rows in the
direction of $\mathbf{F}_L$ are formed, and all vortex rows are
qualitatively equal (all are ``incommensurate''). All rows will
depin approximately at the same driving current. This explains the
sharp $V(I_y)$ transitions for all magnetic fields (see
Fig.~\ref{fig:contourplot}(b)). The integer matching peaks are
still present in $I_{cy}(H)$, since the commensurate vortex
patterns at the integer matching fields $H_n$ can be achieved even
with a small interaction force present, due to a smaller vortex
separation.

Our critical current results are in agreement with what was found
in phase boundary measurements on square SC networks with a
different wire width in the two perpendicular
directions\cite{itzler96physicab}. In these systems, $T_c(H)$
depends on the direction of the current in a similar way as in our
$I_c(H)$ data. This is not surprising since there are many
similarities between SC films with an antidot lattice and SC wire
networks, especially at temperatures close to $T_c$. Using the
temperature to tune the coherence length $\xi(T)$, one can, in
fact observe a weakly coupled network behavior ($\xi(T) \gg$
strips between the antidots) in any antidot array.

Summarizing this part of the discussion, our main experimental
observations can be explained by (i) an \textit{isotropic} single
site pinning force of the rectangular antidots, combined with (ii)
an \textit{anisotropic} vortex-vortex interaction between the
vortices trapped inside the antidots.

\subsection{Numerical simulations}
\label{Sec:pointpinning}

We have used molecular dynamics simulations to confirm that a
model system with the features (i) and (ii) mentioned above, shows
indeed the anisotropy in the critical current that was observed in
the experiments.

\subsubsection{Model}
\label{simulationmodel}

We consider a two-dimensional system with periodic boundary
conditions containing a square array (period $d$) of circular
(i.e. \textit{isotropic}) pinning sites, but with an
\textit{anisotropic} vortex-vortex interaction. The overdamped
equation of motion for a vortex $i$ (Eq.(\ref{eq:overdamped})) is
used to calculate the average vortex velocity as a function of the
applied Lorentz force $\mathbf{F}_L$, and to trace the vortex
trajectories in the pinning array:
\begin{equation}
\eta {{\bf v}_i} = {\bf F}_L + {\bf F}_{vv}({\bf r}_i)
 + {\bf F}_p({\bf r}_i),
\label{eq:overdamped}
\end{equation}
where $\eta$ is the viscosity coefficient and taken to be unity.
The driving force acting on the vortices is the Lorentz force,
${\bf F}_L ={\bf J} \times {\bf \phi}_0$, where $\bf J$ is the
applied current.

The \textit{anisotropic} repulsive vortex-vortex interaction force
between two vortices with separation $r_{ij}$ is described by the
modified Bessel function $K_1$~\cite{Zhu_PP,Zhu_PP2,Brass,Brass2}:
\begin{equation}
 {\bf F}_{vv}({\bf r}_i) = F_{vv0} f_0 \sum_{j \neq i}^{N_v}
 K_1 (\frac {{\bf r}_i - {\bf r}_j}{\lambda}) {\bf r}_{ij},
\label{eq:vvinteraction}
\end{equation}
where
$\mathbf{r}_{ij}=\frac{(\mathbf{r}_i-\mathbf{r}_j)}{|\mathbf{r}_i-\mathbf{r}_j|}$,
and $N$ denotes the number of vortices in the sample. $F_{vv0}f_0$
expresses the intensity of the vortex-vortex interaction force,
with $f_0=\frac{\phi^2}{8 \pi^2 \lambda^3}$. The cut-off for the
vortex-vortex interaction lies at $6~d$ in both the $x$- and the
$y$-direction. The vortex-vortex interaction is made anisotropic
by choosing $\lambda_y= 2 \lambda_x=d$. All forces are expressed
in units of $f_0$.

The rectangular antidots of the experiment are modeled as
potential wells with an \textit{isotropic} attractive pinning
force $\mathbf{F}_p$ given by:
\begin{equation}
{\bf F}_p({\bf r}) = - F_{p0} f_0 \sum_{k=1}^{N_p} \frac{\bf
r}{R_p} exp\left( - \left\vert \frac{\bf r}{R_p} \right\vert^2
\right) ,
\end{equation}
Here, ${\bf r}$ is the distance between the vortex and the
$k^{th}$ pinning site.

Calculations are performed on a square sample (side $8~d$) with
periodic boundary conditions in both $x$- and $y$-directions. The
same initial vortex configuration, obtained from an annealing
course~\cite{reichhardt0800losalamos}, is used for the hysteretic
calculation of the $\langle v \rangle(F_L)$ current-voltage
characteristics along the $x$- and the $y$-direction.

We solve the equation of motion (Eq.(\ref{eq:overdamped})) by
taking discrete time steps $\Delta t$, using the following fixed
lengths and forces: $R_p = 0.7$, $\lambda_y = 6$, $F_{vv0} = 0.2$,
$F_{p0} = 1$. The average vortex velocity as a function of applied
driving force $\langle v \rangle(F_L)$ is obtained by increasing
$F_L$ in small steps $\Delta F_L = 0.001$, starting from $F_L =0$.
At each $F_L$, we neglect the first 2000 time steps, and average
the next 8000 steps to calculate the average velocity of all the
vortices in the system.

\subsubsection{Results}

\begin{figure}[hbc]
  \centering
  \includegraphics*[scale=0.7]{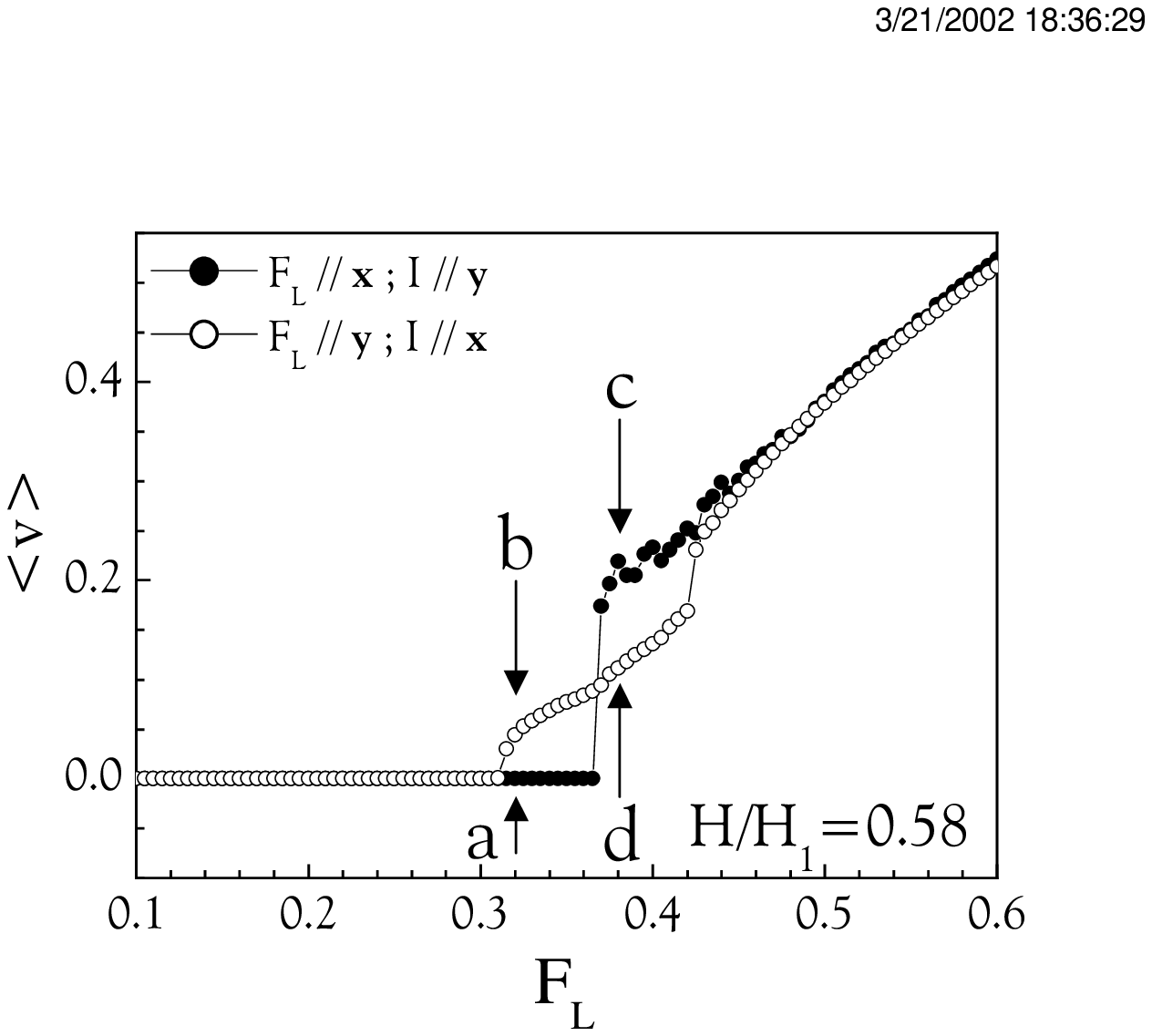}
  \caption{Average vortex velocity as a function of the
  Lorentz force $F_L$ at $H/H_1=0.58$. The open (filled) symbols show the result
  for a current along the $x$- ($y$-) axis. Arrows $a$ and $c$ ($b$ and $d$)
  indicate the points at which the vortex trajectories for $I~\|~\mathbf{y}$
   ($I~\|~\mathbf{x}$) were calculated, shown in Fig.~\protect\ref{fig:trajectories}.}
  \label{fig:v(i)simulation}
\end{figure}

The obtained average vortex velocity as a function of the applied
Lorentz force for $H/H_1=0.58$, i.e. a magnetic field in between
matching fields, is shown in Fig.~\ref{fig:v(i)simulation}.  We
will compare these calculated $\langle v \rangle(F_L)$ curves at
$H/H_1=0.58$ with the experimental $V(I)$ curves taken at
$H/H_1=0.4$ (Fig.~\ref{fig:iv}), which are typical for all
magnetic fields in between matching fields. The two main features
observed in the experimental curves are qualitatively reproduced.
First, the onset of the vortex motion occurs at lower driving
force for $I~\|~\mathbf{x}$ ($\mathbf{F}_L~\|~\mathbf{y})$ (open
symbols) than for $I~\|~\mathbf{y}$ ($\mathbf{F}_L~\|~\mathbf{x})$
(filled symbols). The critical depinning force in the
$y$-direction lies at $F_L \approx 0.315$, while it is $F_L
\approx 0.37$ when the force is applied in the $x$-direction.
Therefore, the critical current for $I~\|~\mathbf{x}$ is lower
than for $I~\|~\mathbf{y}$. Secondly, the $V(I)$ transition is
clearly much broader for $I~\|~\mathbf{x}$ than for
$I~\|~\mathbf{y}$, with a gradually increasing average vortex
velocity.

The origin of the different $\langle v \rangle (F_L)$ behavior for
the two directions of the Lorentz force can be found by examining
the trajectories of the depinned vortices
(Fig.~\ref{fig:trajectories}) at two Lorentz force values,
indicated by arrows in Fig.~\ref{fig:v(i)simulation}. A Lorentz
force of 0.32, applied in the $x$-direction (arrow $a$ in
Fig.~\ref{fig:v(i)simulation}) is not sufficient to depin any
vortex rows (see Fig.~\ref{fig:trajectories}(a)). When the same
$F_L=0.32$ is applied in the $y$-direction (arrow $b$ in
Fig.~\ref{fig:v(i)simulation}), a \textit{partial depinning} takes
place, where the rows with little order in the $y$-direction
(incommensurate rows) are the first ones to start moving (see
Fig.~\ref{fig:trajectories}(b)). The commensurate rows (the first,
second, and forth row, counting from the left) remain pinned (see
also Section~\ref{sec:exp_discussion}). At a Lorentz force
$F_L=0.38$, applied along the $x$-direction (arrow $c$ in
Fig.~\ref{fig:v(i)simulation}), all the vortex rows have started
to move (Fig.~\ref{fig:trajectories}(c)). This corresponds to the
sharp increase in $\langle v \rangle$ that can be found in the
$\langle v \rangle(F_L)$ curve (filled symbols in
Fig.~\ref{fig:v(i)simulation}). For the same $F_L=0.38$ applied in
the $y$-direction (arrow $d$ in Fig.~\ref{fig:v(i)simulation}),
the vortices in the incommensurate rows move with a higher
velocity (Fig.~\ref{fig:trajectories}(d)). When an even higher
Lorentz force is applied (e.g. $F_L=0.44$), all vortices finally
move for both $\mathbf{F}_L$ directions (not shown).

\subsubsection{Discussion}
The results of these molecular dynamics simulations show that an
array of \textit{isotropic pinning sites}, combined with an
\textit{anisotropic vortex-vortex interaction} along the two
principal axes of the PPA, gives rise to the same phenomena as
observed in our pinning experiments on a Pb film with a square
array of rectangular antidots. The low critical current and broad
$V(I)$ transitions in between matching fields for
$I~\|~\mathbf{x}$ ($\mathbf{F}_L~\|~\mathbf{y}$) are found to be
due to the motion of the incommensurate vortex rows along the
direction of the strong vortex-vortex interaction, in this case
the $y$-direction. For $I~\|~\mathbf{y}$
($\mathbf{F}_L~\|~\mathbf{x}$), the simultaneous vortex depinning
at a relatively high driving force is responsible for the high
critical current and the sharp $V(I)$ transitions.  This result
strengthens our interpretation that in our experiment, the
rectangular antidots provide an \textit{isotropic pinning
potential} but induce an \textit{anisotropy in the vortex-vortex
interaction}.
\begin{figure}[hct]
  \centering
  \includegraphics*[scale=0.80]{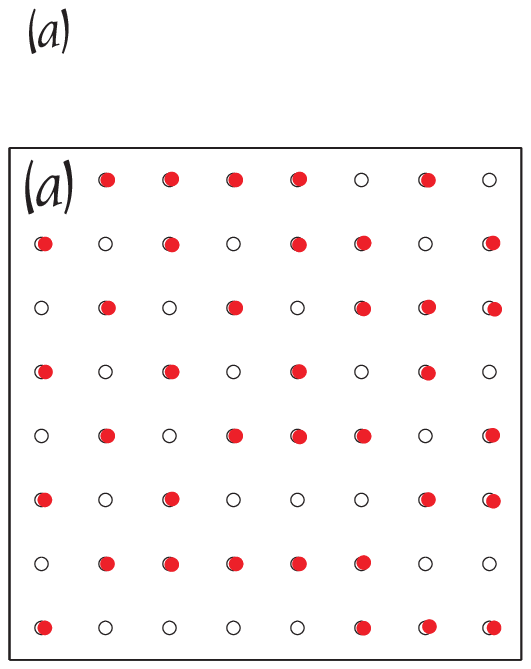}
  \includegraphics*[scale=0.80]{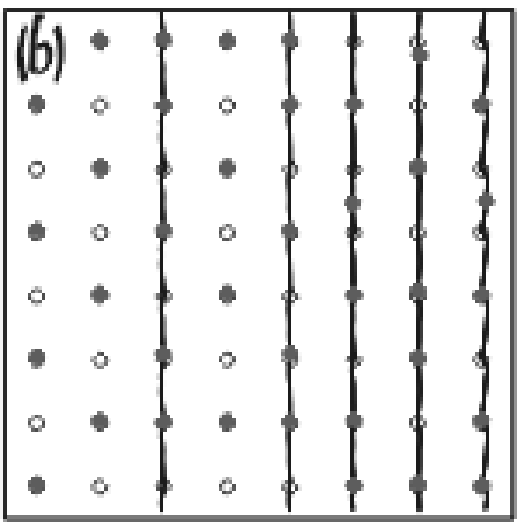}
  \includegraphics*[scale=0.80]{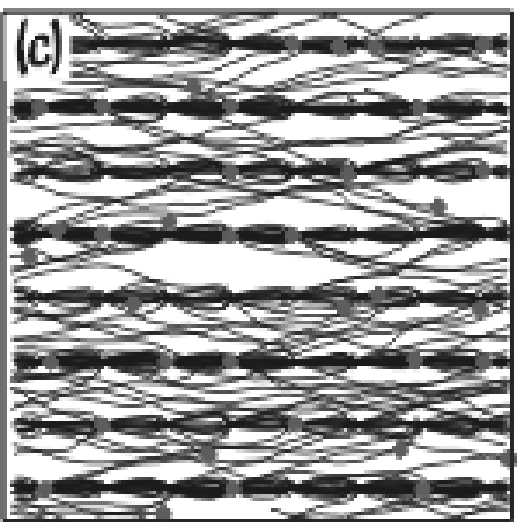}
  \includegraphics*[scale=0.80]{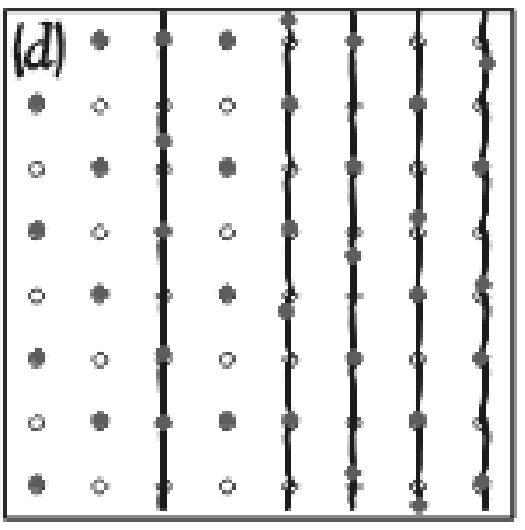}
  \caption{Vortex trajectories (lines) for $H/H_1=0.58$ at $F_L=0.32$
  for a current in the $y$- (a) and the $x$-direction (b). Open
  circles represent pinning sites, black dots represent vortices.
  Idem at $F_L=0.38$ for the current in the $y$- (c) and
  $x$-direction (d). The labels (a) to (d) correspond to the arrows in
  Fig.~\protect\ref{fig:v(i)simulation}.}
  \label{fig:trajectories}
\end{figure}

Due to the anisotropic nature of the vortex-vortex interaction,
the initial vortex lattice found in the simulation by an annealing
course shows more order in the $y$-direction (stronger
vortex-vortex interaction) than in the $x$-direction (weaker
vortex-vortex interaction) (see Fig.~\ref{fig:trajectories}(a)).
Although we believe that this is also what happens in the
experiment, no definitive statement about the experimental vortex
configuration can be made, however, at this stage, before a direct
experimental study of the vortex patterns in these anisotropic
pinning arrays, e.g. by scanning Hall
probe\cite{bending00physicac, field00losalamos} or
Lorentz\cite{harada96science} microscopy.

It is interesting to compare our results with the molecular
dynamics simulations on a \textit{rectangular} array of isotropic
pinning sites, recently performed by Reichhardt \textit{et al.}
\cite{reichhardt1100losalamos}. In that case, the anisotropy in
$\mathbf{F}_{VV}$ between pinned vortices is induced by a
different spacing in the $x$- and $y$-direction of the pinning
sites. A similar anisotropy in $I_c(H)$ as in our measurements and
calculations is indeed found when no interstitial vortices are
present ($H<H_1$). When the Lorentz force $\mathbf{F}_L$ is in the
direction of the weak vortex-vortex interaction (in the case
considered in Ref. \cite{reichhardt1100losalamos} along the
direction with a large lattice spacing) the absence of rational
matching and an overall high $I_c(H)$ is found.

\subsection{Summary}

We have measured $I_c(H)$ and $V(I,H)$ curves of a SC film with a
square array of rectangular antidots for two directions of the
applied current. We find an \textit{overall high} $I_c(H)$ with
integer matching peaks but no rational matching features when the
current is applied parallel to the long side of the antidots. This
is a clear advantage compared to isotropic pinning arrays, where a
large suppression of $I_c(H)$ is seen in between the rational
matching fields. We attribute this effect to the
\textit{anisotropic vortex-vortex interaction}, which is stronger
along the long side of the antidots. Molecular dynamics
simulations on a square array of \textit{isotropic} pinning sites,
combined with an \textit{anisotropic} vortex-vortex interaction,
indeed show the same characteristic anisotropic features as
observed in our measurements.


\section{Square array of rectangular blind antidots: Numerical simulations}

In this Section, we examine the vortex mobility in a
superconductor with a square array of rectangular \textit{blind}
antidots by means of molecular dynamics simulations. Blind
antidots are holes that do not perforate the superconductor
completely, i.e. the bottom of a blind antidot always contains
superconducting material. Therefore, if more than one vortex is
trapped inside a blind antidot, the vortices remain individual
repulsive entities, without merging into one multiple-quantum
vortex~\cite{vvm96prb,bezryadin96}, as in the case of an antidot.

Moreover, a vortex trapped inside a blind antidot can move around
freely within the boundaries of the pinning center. In a film with
antidots, on the other hand, the supercurrents constituting the
vortex are forced to flow at the outer edge of the antidot,
confining the vortex in space.

\subsection{Model}

Taking into account the properties of a blind antidot, mentioned
above, we will use the following model to simulate the vortex
motion in a superconductor with an array of blind antidots.

Again, we use a square 2D system (size $8d$) with periodic
boundary conditions in the $x$- and $y$-direction. The system
contains a square array (period $d$) of rectangular pinning sites
(sides $R_x=0.4d$ and $R_y=0.77d$). The vortex velocities are
calculated using the normalized overdamped equation of motion
Eq.~(\ref{eq:overdamped}). The repulsive vortex-vortex interaction
is described by Eq.~(\ref{eq:vvinteraction}), using an
\textit{isotropic} penetration depth $\lambda=d$ and $F_{vv0} =
0.2$.

\begin{figure}[ht]
  \centering
  \includegraphics*[scale=0.45]{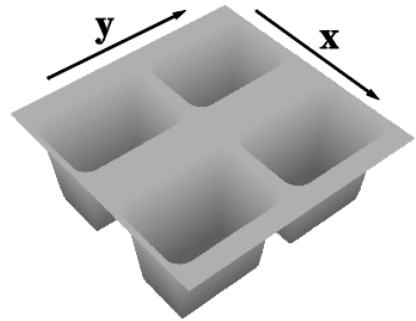}
  \caption{3D plot of the pinning potential used to model a square array
of rectangular blind antidots.}
  \label{fig:fig1a-b}
\end{figure}

The pinning force of a rectangular pinning center on the vortex
can be defined as~\cite{Zhu_PP}:
\begin{equation}
{\bf F}_p({\bf r}) = - F_{p0} f_0 \sum_{k=1}^{N_p} \frac{\bf
r}{R_p} exp\left( - \left\vert \frac{\bf r}{R_p} \right\vert^2
\right) ,
\end{equation} when the vortex is outside the $k^{th}$ rectangular
pinning center, and ${\bf F}_p({\bf r}) \equiv 0$, if the vortex
is inside the rectangular area of the pinning site. This means
that no pinning force is exerted on the vortex, once it is trapped
inside the blind antidot. In this expression, ${\bf r}$ is the
distance between the vortex and the edge of the $k^{th}$
rectangular pinning center. We have chosen $R_p$ to be $R_p =
0.067 d $ and $F_{p0} = 1$. In Fig.~\ref{fig:fig1a-b}, we show
schematically the 3D pinning potential that was used for this
simulation.

The current-voltage characteristics $\langle v \rangle (F_L)$ and
the vortex trajectories for a current along the $x$- and the
$y$-direction are calculated using the method described in
Section~\ref{simulationmodel}.

\subsection{Results and discussion}

\begin{figure}[ht]
  \centering
  \includegraphics*[scale=0.7]{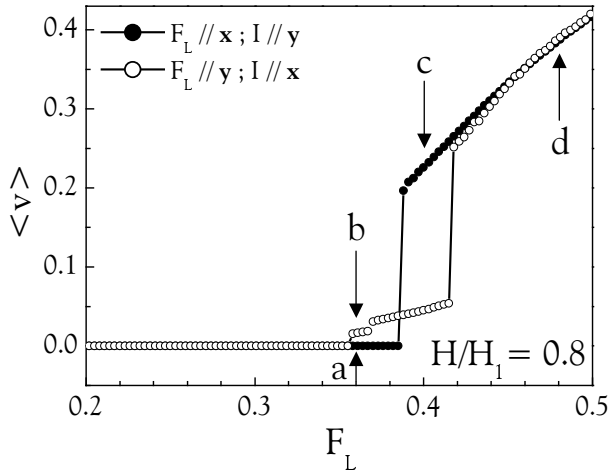}
  \caption{Average vortex velocity as a function of the Lorentz force
   ${\bf F}_L$ at $H/H_1=0.8$. The open (filled) symbols show the result
   for a current along the $x$- ($y$-) axis. Arrows $a$ and $c$ ($b$ and
   $d$) indicate the points at which the vortex trajectories for
   $I~\|~\mathbf{y}$  ($I~\|~\mathbf{x}$) were calculated, shown in
   Fig.~\protect \ref{fig:fig3a-d}.}
  \label{fig:zhu_fig2}
\end{figure}

In Fig.~\ref{fig:zhu_fig2}, we show the calculated $\langle v
\rangle(F_L)$ curves at $H/H_1 =0.8$, for a driving force ${\bf
F}_L$ in the $x$- (filled symbols) and the $y$-direction (open
symbols). The magnetic field is chosen in between matching fields,
where no commensurate vortex configuration can be established.

The calculation shows that the onset of vortex motion occurs at a
lower Lorentz force $F_L$ when $\mathbf{F}_L~ \|~\mathbf{y}$. For
this direction of the driving force, the average vortex velocity
becomes finite at $\mathbf{F}_L \approx 0.356$. When the Lorentz
force is applied in the perpendicular direction ($\mathbf{F}_L~\|
~\mathbf{x}$), vortex motion is delayed until $\mathbf{F}_L
\approx 0.386$. Consequently, the critical current $I_c$ is
smaller for $I ~ \| ~\bf{x}$ than for $I~ \| ~ \bf{y}$.

Moreover, there is a striking difference in the width of the
depinning transition. When $\mathbf{F}_L~\|~\mathbf{y}$, we
observe several steps in the depinning process, leading to a broad
transition, while for $\mathbf{F}_L~ \| ~\mathbf{x}$, the $\langle
v \rangle(F_L)$ curve shows a single sharp jump at $F_L \sim
0.386$.

\begin{figure}[b]
  \centering
  \includegraphics*[scale=0.75]{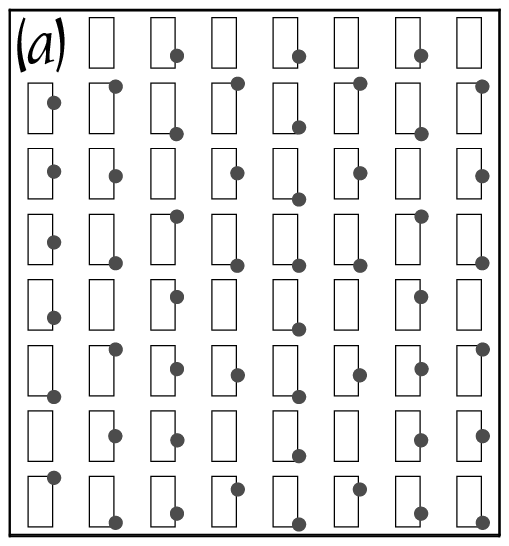}
  \includegraphics*[scale=0.75]{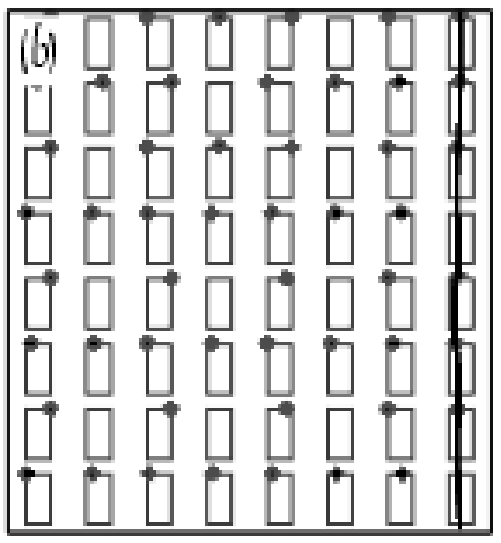}
  \includegraphics*[scale=0.75]{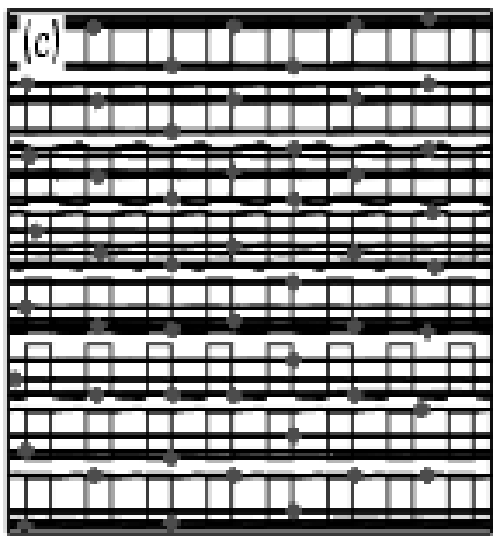}
  \includegraphics*[scale=0.75]{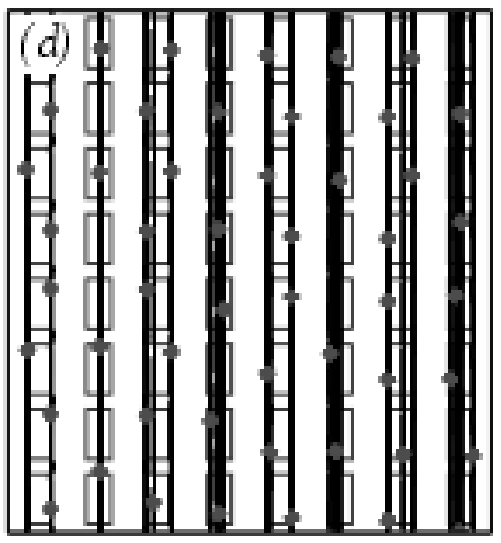}
  \caption{Vortex trajectories (lines) for $H/H_1=0.8$
  at $F_L = 0.36$ for a current in the $y$- (a) and $x$-direction (b).
  Rectangles represent the blind rectangular antidots, black dots
  represent vortices. Idem at $F_L=0.40$ for a current in the
  $y$-direction (c), and at $F_L=0.48$ for a current in the
  $x$-direction. The labels (a) to (d) correspond to the arrows in
  Fig.~\protect\ref{fig:zhu_fig2}.}
  \label{fig:fig3a-d}
\end{figure}

We examine the origin of this qualitatively different behavior for
the two directions of the Lorentz force by plotting the vortex
trajectories and positions after 10$^5$ time steps (see
Fig.~\ref{fig:fig3a-d}) at some specific Lorentz force values,
indicated by arrows in Fig.~\ref{fig:zhu_fig2}.

First, we observe a different configuration of the pinned vortices
in Fig.~\ref{fig:fig3a-d}(a) and (b), although they originate from
the same annealing course. The difference results from vortex
jumps that occur after the annealing course, under the influence
of the applied Lorentz force. These jumps aid the system to obtain
its lowest energy configuration. At a magnetic field of
$H/H_1=0.8$, it is not possible to obtain a completely regular
vortex pattern. The commensurate configuration, which is closest
to $0.8H_1$ is $0.75 H_1=H_{3/4}$, where one finds consecutively
completely filled rows and rows with half-filling. At $H/H_1=0.8$,
the energetically most favorable vortex configuration looks like
the configuration at $H_{3/4}$, but with some excess vortices that
are placed in the rows with half-filling. Because the jumps are
assisted by the Lorentz force, we find this kind of order in the
vortex positions only in the direction of the applied Lorentz
force (along the $x$-direction in Fig.~\ref{fig:fig3a-d}(a) and
along the $y$-direction in Fig.~\ref{fig:fig3a-d}(b)).

A Lorentz force of 0.36 applied in the $x$-direction (arrow $a$ in
Fig.~\ref{fig:zhu_fig2} is not sufficient to depin any vortex rows
(see Fig.~\ref{fig:fig3a-d}(a)). When the same $F_L=0.36$ is
applied in the $y$-direction (arrow $b$ in
Fig.~\ref{fig:zhu_fig2}), a \textit{partial depinning} takes
place, where the row with the least order in the $y$-direction
(the first vertical row counting from the right in
Fig.~\ref{fig:fig3a-d}(b)), is set in motion. The moving vortex
row is the only one that contains 6 vortices. Accommodating these
six vortices in a row of eight pinning sites gives rise to a lot
of strain, making it considerably easier to depin this
incommensurate row. Most of the other rows contain either eight or
four vortices, which can be easily positioned in a low energy
configuration (commensurate rows). One vertical row accommodates
five vortices. This row will be the next to be depinned when the
Lorentz force is increased (at $F_L \approx 0.37$), creating the
second step in the $\langle v \rangle (F_L)$ curve for
$\mathbf{F}_L~\|~\mathbf{y}$ in Fig.~\ref{fig:zhu_fig2} (open
symbols).

At a Lorentz force of $F_L=0.4$, applied along the $x$-direction,
all vortices are set in motion (Fig.~\ref{fig:fig3a-d}(c)). This
corresponds to the sharp increase in $\langle v \rangle $ that is
seen in the $\langle v \rangle (F_L)$ curve (filled symbols in
Fig.~\ref{fig:zhu_fig2}). When a Lorentz force of $F_L=0.48$ is
applied in the $y$-direction (arrow $d$ in
Fig.~\ref{fig:zhu_fig2}), all vortices participate in the motion.

In Fig.~\ref{fig:fig3a-d}(a), one also finds commensurate
horizontal rows, containing eight or four vortices, one
incommensurate horizontal row with five and one with six vortices.
However, these incommensurate rows do not give rise to a partial
depinning as in the case $\mathbf{F}_L~\|~\mathbf{y}$, because the
strain that is caused by the incommensurability is effectively
released by \textit{shifting the vortex positions within the
rectangular blind antidot}. When $\mathbf{F}_L~\|~\mathbf{x}$, the
long side ($R_y$) of the rectangular blind antidot can be used for
this purpose. For $\mathbf{F}_L~\|~\mathbf{y}$, the vortices are
all located at the top edge of the rectangular antidots.
Consequently, only the short side of the antidots ($R_x$) is
available, resulting in large stresses in the incommensurate
vertical rows which initiate the partial depinning proces.

We also notice that there is some amount of row switching at high
driving force for $\mathbf{F}_L~\|~\mathbf{x}$ due to the smaller
spacing between the blind antidots in that direction. Although the
plotted trajectory lines in Fig.~\ref{fig:fig3a-d}(c) do not show
it directly, we know the switching takes place by comparing the
number of vortices in a particular horizontal row at low and at
high driving force. On the other hand, \textit{no} row switching
is seen when $\mathbf{F}_L~\|~\mathbf{y}$.

\begin{figure}[hct]
  \centering
  \includegraphics*[scale=0.7]{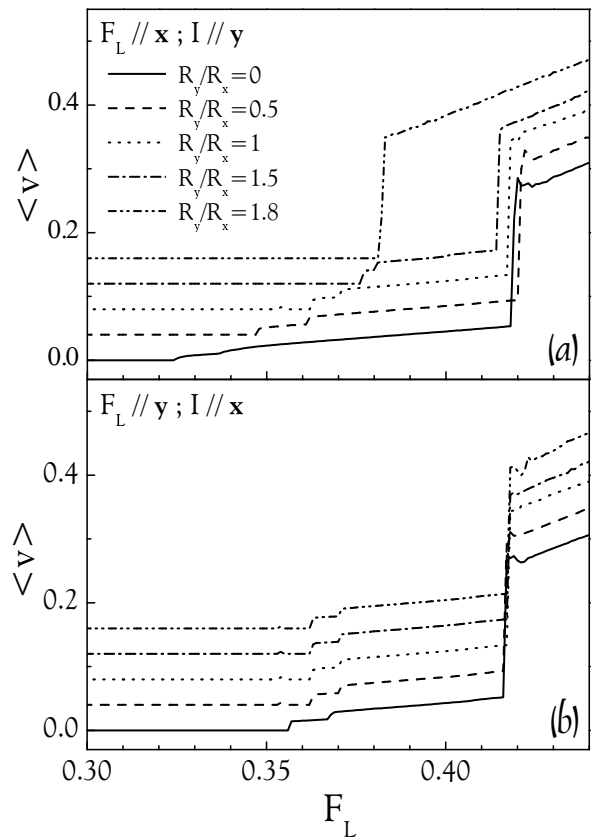}
  \caption{Average vortex velocity as a function of the Lorentz force
   $\mathbf{F}_L$ at $H/H_1=0.8$ with $\mathbf{F}_L ~\| ~\mathbf{x}$
   (a) and $\mathbf{F}_L ~\| ~\mathbf{y}$ (b). $R_x = 0.4
   d$ is kept fixed and the ratio $R_y/R_x$ is varied from 0.0 to 1.8, as indicated.}
  \label{fig:fig8}
\end{figure}

We have also examined the effect of the side ratio of the
rectangular blind antidots $R_y / R_x$ on the vortex depinning. In
Fig.~\ref{fig:fig8}, we present $\langle v \rangle (F_L)$ curves
(shifted vertically for clarity) at $H/H_1 = 0.8$ for various
$R_y/R_x$, smaller than the ratio used for
Fig.~\ref{fig:zhu_fig2}.

When $\mathbf{F}_L~\|~\mathbf{y}$ (Fig.~\ref{fig:fig8}(b)), the
$\langle v \rangle (F_L)$ curves are almost independent on the
ratio $R_y/R_x$. The transition is always a two-step process,
involving partial depinning of the incommensurate vertical rows,
as described above.

When $\mathbf{F}_L~\|~\mathbf{x}$ (Fig.~\ref{fig:fig8}(a)), we
find a sharp depinning transition, resulting from simultaneous
depinning of all vortices and a high critical current, if $R_y$ is
sufficiently large, $R_y \geq 1.8$ (see also
Fig.~\ref{fig:zhu_fig2}, where $R_y/R_x=1.92$). For $R_y/R_x \leq
1.5$, two step transitions in the $\langle v \rangle (F_L)$ curves
are always seen. The onset of vortex motion also appears to be
strongly dependent on the ratio $R_y/R_x$. As long as $R_y > R_x$,
the critical current is higher when $\mathbf{F}_L~\|~\mathbf{x}$.
When $R_y < R_x$, the highest critical current is found for
$\mathbf{F}_L~\|~\mathbf{y}$.

When $\mathbf{F}_L~ \|~ \mathbf{y}$, the important length scale
determining the confinement of the trapped vortices is the width
$R_x$ of the rectangular blind antidots. Since $R_x$ is kept
constant in the calculations, we indeed expect to find no
dependence of the $\langle v \rangle (F_L)$ transitions on
$R_y/R_x$.

When $\mathbf{F}_L~ \|~ \mathbf{x}$, on the other hand, the
initial critical depinning force is very sensitive to $R_y$,
because now this parameter determines the confinement of the
vortices. When $R_y/R_x = 0$, i.e. $R_y=0$, the pinning centers
are in fact points in the $x$-direction, which results in a very
low threshold value for the initial vortex depinning transition.
When $R_y/R_x$ increases, the vortices trapped in the rectangular
blind antidots obtain more freedom to shift their positions along
the $y$-axis. This gradually improves their ability to reduce the
strain arising from the vortex-vortex interaction in an
incommensurate row. In this way, the initial vortex depinning of
the incommensurate rows can be delayed to higher $F_L$.

\subsection{Summary}

We have investigated numerically vortex pinning and dynamics in a
square array of rectangular blind antidots for different
orientations of the applied current or Lorentz force. We
demonstrate that the onset of vortex motion, and therefore the
critical current, depends on the orientation of the Lorentz force.
The critical current is found to be higher when the current is
applied parallel to the long side of the antidots. By shifting
their position inside the blind antidot, the vortices are able to
delay the depinning of the incommensurate vortex rows to higher
$F_L$, by reducing the strain arising from the vortex-vortex
interaction in the incommensurate rows. We have also found a
strong dependence of the $\langle v \rangle (F_L)$ transition on
the aspect ratio of the rectangular blind antidots.

\section{Comparison of the antidot and the blind antidot system}

This paper presents results on the anisotropy in vortex pinning
for two very different systems: a superconducting film with a
square array of rectangular antidots, and with a square array of
rectangular \textit{blind} antidots. We have shown that in both
systems the critical current is higher and the IV-transition is
sharper when the current is sent parallel to the long side of the
antidots. On the other hand, a current along the short side of the
rectangular antidots yields a lower critical current and a broad
IV-transition in both systems.

Although the behavior of the antidot and blind antidot system
appears to be similar, the mechanisms determining this behavior
are very different in the two systems. In the case of the
superconducting film with rectangular \textit{antidots}, the
experiments show that the pinning force of the rectangular
antidots is identical for the two perpendicular directions. Here,
the \textit{vortex-vortex interaction} is found to be anisotropic
and responsible for the observed anisotropy in the critical
current measurements. In the case of rectangular \textit{blind
antidots}, the vortex-vortex interaction is isotropic. Here, the
anisotropy in the shape of the antidots and the fact that a vortex
can move around freely within the blind antidot cause the
anisotropy in the critical current.

\section{Conclusion}

We have investigated vortex pinning and dynamics in
superconducting films with a square array of rectangular antidots
and blind antidots. As shown by experiments (on the antidots
system) and numerical calculations (on both systems), the critical
current is higher when applied parallel to the long side of the
antidots. For this current direction, we also find sharper IV
transitions in both systems. By examining the vortex trajectories
after depinning by means of numerical simulations, we were able to
demonstrate that the origin of the anisotropy is very different in
the antidot and the blind antidot system. Nevertheless, both
systems illustrate how patterning of an otherwise isotropic
superconducting film, can induce an important anisotropy in the
critical current and IV characteristics.

\section*{Acknowledgements}
This work was supported by the ESF "Vortex" Program, the bilateral
BIL 00/02 China/Flanders Project, the Belgian Interuniversity
Attraction Poles (IUAP), the Flemish GOA and FWO Programs, and the
National Natural Science Foundation of China, the Ministry of
Science and Technology of China (NKBRSF-G1999064602). We wish to
thank M.~J.~Van Bael, Y.~Bruynseraede, C.~Reichhardt, C.~J.~Olson
and F.~Nori for helpful discussions.

\end{document}